\documentclass[12pt,preprint]{aastex}
\usepackage{amsmath, amsthm, amssymb}

  \def\ltsima{$\; \buildrel  <  \over \sim
  \;$}    \def\simlt{\lower.5ex\hbox{\ltsima}}    
\def\gtsima{$\;      \buildrel      >      \over      \sim      \;$}
\def\simgt{\lower.5ex\hbox{\gtsima}}      

\newcommand{\nox}{N_{\rm O}^{\rm X}}
\newcommand{\nou}{N_{\rm O}^{\rm UV}}
\newcommand{\mnhi}{N_{\rm HI}}
\def\cm#1{\, {\rm cm^{#1}}}
\def\lya{Ly$\alpha$}
\newcommand{\rcbm}{r_{\rm CBM}}
\newcommand{\rhion}{r_{\rm HI}}
\newcommand{\roion}{r_{\rm OIX}}
\newcommand{\rpre}{r_{\rm preion}}
\newcommand{\rstep}{r_{\rm step}}

\begin{document}

   
\title{An Explanation for the Different X-ray to Optical Column
  Densities in the Environments of Gamma Ray Bursts: A
  Progenitor Embedded in a Dense Medium}
\author{Yair Krongold\altaffilmark{1} and  J. Xavier Prochaska\altaffilmark{2}}
\altaffiltext{1}{Instituto de Astronomia, Universidad Nacional
  Autonoma de Mexico, Apartado Postal 70-264, 04510 Mexico DF,
  Mexico.}  
\altaffiltext{2}{Department of Astronomy and Astrophysics, 
  UCO/Lick Observatory; University of California, 1156 High Street, Santa Cruz, 
  CA 95064; xavier@ucolick.org}


\begin{abstract} 
We study the $\gtrsim10$ ratios in the X-ray to optical
column densities inferred from afterglow spectra of Gamma Ray Bursts
due to gas surrounding their progenitors. We
present time-evolving photoionization calculations for these
afterglows and explore different conditions for their
environment. We find that homogenous models of
the environment (constant density) predict X-ray columns similar to
those found in the optical spectra, with the bulk of the opacity being
produced by neutral material at large distances from the burst. This
result is independent of gas density or metallicity. 
Only models assuming a progenitor immersed in a dense ($\sim10^{2-4}$ cm$^{-3}$) cloud of gas (with radius $\sim10$ pc),  with a strong, declining gradient of
density for the surrounding interstellar medium are able to account
for the large X-ray to optical  column density ratios. However, to avoid an unphysical correlation between the size of this cloud, and the size of the ionization front produced by the GRB, the models also require that the circumburst medium is already ionized prior to the burst.
The inferred cloud masses are $\lesssim10^6$ M$_\odot$, even if low metallicities in the medium are assumed ($Z\sim 0.1~Z_\odot$). These cloud properties are
consistent with those found in giant molecular clouds and our
results support a scenario in which the progenitors  reside within
intense star formation regions  of galaxies. Finally, we show that
modeling over large samples of GRB afterglows may
offer strong constraints on the range of
properties in these clouds, and the host galaxy ISM. 
\end{abstract}


\section{Introduction \label{par:intro}}


GRBs are the most powerful sources of radiation in the
Universe. Although they shine for brief periods of time, their
afterglows are capable of ionizing their surrounding material to large
distances --  tens to hundreds of pc
\citep{pl98a,pl02,pdr+08,pwf+08,dfp+09}.
Therefore, GRB afterglows can
ionize the  medium close to the progenitor, the region of star
formation containing the progenitor, and even part of the
inter-stellar medium (ISM) of the host galaxy. Furthermore, as the
light travels along our line of sight, GRB afterglows ``illuminate''
the host galaxy in absorption giving us a detailed view of the
surrounding gas.  As a result, GRBs
have been used over the last decade as probes of the physical
conditions of the ISM in distant galaxies \citep[e.g.][]{savaglio06,pcd+07}.
In addition, it has been established that GRBs probe
different regions than Quasars (likely because GRBs originate in star-forming
regions), making studies of the environments of both objects
complementary to one another \citep{vel+04,pcd+07}.

The material around GRBs has been observed spectroscopically both 
at X-ray and optical/UV frequencies. There is    
clear evidence for intrinsic absorption due to high columns of
material in the X-ray spectra for the majority of GRB afterglows
\citep{bph+06,whf+07,csm+12}.
The measured opacities imply metal column densities gauged by oxygen
of $\nox \approx 10^{19} \cm{-2}$. 
For a solar metallicity, which is likely rare
\citep{pcd+07,srg+12}, this
implies an effective hydrogen column density $N_{\rm H} \approx
10^{22} \cm{-2}$.  
Optical/UV spectroscopy also reveals strong 
intrinsic absorption in a large fraction ($\approx 90\%$) 
of sightlines to GRBs \citep{sff03,pcb+07,fjp+09}.
From these data, one may estimate the column densities for oxygen 
from the measurements of unsaturated transitions of Si, S, and Zn
assuming solar relative abundances.   Typical values are
$\nou \approx 10^{18} \cm{-2}$ with a large dispersion.
One also measures, from damped \ion{H}{1} \lya\ absorption, neutral
hydrogen column densities of $\mnhi \approx 10^{21-22}  \cm{-2}$ \citep{jfl+06}. 
These X-ray and optical/UV spectroscopic observations, each require
a large reservoir of gas along the sightline, presumably tracing the
circumburst medium and the larger-scale, ambient ISM of the galaxy.

However, when a direct comparison between  the X-ray and optical
inferred column densities toward individual GRBs has been possible, a
discrepancy has been observed:  the inferred column density of atoms
and ions from X-ray studies are systematically larger (by
factors of a few to several orders of magnitude) than the column
densities measured in the optical/UV 
\citep{whf+07,ssk+11,csm+12}\footnote{Standard treatment
  in the literature is to recast these metal column densities as
  equivalent H column densities, but that is not necessary nor
  preferred for the following analysis.}.   
Since the absorption observed in the X-rays is
sensitive to the total column density of material in the line of sight
(produced from neutral atoms up to highly ionized, but not fully
stripped species), while that in the optical/UV region depends on the
ionization state of the gas (probing better absorption by cooler gas),
it has been proposed that the discrepancies are the result of
photoionization  of the gas close to the GRB by its own radiation field
\citep{whf+07,ssk+11}.
The large X-ray column densities also suggest a
dense environment near the burst location \citep[][]{csm+12} while analysis of fine-structure transitions in the
optical/UV indicate the neutral gas lies at distances of $\approx
100$\,pc to many kpc  \citep{pcb06,dcp+06,vls+07,dfp+09,sheffer+09}.

While these are plausible hypotheses, quantitative work is required
to explore the viability and implications for the circumburst
medium and beyond.  
The principal challenge is to construct a physically realistic model
with sufficient opacity in the ionized gas on small scales without
over-predicting the observed column densities of lower ionization
state gas. If the medium has roughly a constant density, the column density of the colder gas can easily supersede that of the ionized material even at  low distances from the burst (see \S \ref{sec:disc}), suggesting the presence of an inhomogeneous distribution of material around the burst (with denser gas closer to the burst region). 

In order to study the circumburst medium around GRBs, considering in a
self-consistent manner both the gas distribution and its ionization
structure, one must consider that the gas is far from ionization
equilibrium after the onset of the burst (\S \ref{sec:tei}).  As has
been shown by  calculations carried out in the past, following the
time evolution of the ionization state of the gas is required to
obtain reliable models of the surrounding material.  This has been
amply demonstrated by \citet[][]{lp+02}, \citet[][]{pl+02}, \citet[][]{plf+03}, and \citet[][]{lp+03}. These authors
present observable constraints on the X-ray versus Optical (dust
continuum) extinction, based on self-consistent models considering
time evolving calculations for both the dust fraction and the
ionization state of the gaseous phase. Special attention is put on the
effects of dust on the medium. More recently, \citet[][]{pdr+08}
have studied the time effects on the ionization structure and front in
the circumburst medium \citep[see also][]{pwf+08,dfp+09,dfp2+09}. These studies have been valuable in determining the
effects of the GRB in their medium, however, they have not attempted
to explain the different column densities explained above, and have
assumed homogeneous gas distributions.

As a first step to further explore these differences, 
we have used a custom time-dependent photoionization code 
to predict the relative column densities of gas relevant to X-ray and
optical/UV observations.  The code computes the evolution of the different ionization species as a function of time and distance, given the luminosity, light curve, and spectral energy distribution of a source emitting  ionizing photons. 
To focus the discussion, we compare results obtained so far in the
literature against a fiducial model representative of the 
medium surrounding GRBs.  These tend to show
an excess of $\sim$10 in column
density between the X-ray  and optical spectra of the
source.

The paper is organized as follows: In \S \ref{sec:tei} we describe briefly the code and provide a test case to exemplify in a simple way the time evolution of the circumburst medium. In \S \ref{sec:fiducial} we lay out the  assumptions for the particular fiducial models presented here. In \S \ref{sec:disc} we present and analyze our results. Finally,  \S \ref{sec:conc} summarizes our findings. 
Throughout the paper we have assumed a Hubble constant
H$_0$=72 km s$^{-1}$ Mpc$^{-1}$, a dark energy fraction $\Omega_\Lambda$=0.75, and a matter fraction $\Omega_m$=0.25.

\section{Time-Evolving Photoionization Calculations \label{sec:tei}}

GRB afterglows have roughly a power-law spectrum (due to synchrotron
processes),  with emission extending from X-ray to radio frequencies. 
Therefore, the emission of GRB afterglows contains a large
fraction of ionizing photons that will ionize and heat the surrounding
material. These photons will produce an ionization front, that will
expand over time as additional radiation impinges on the gas. An ionization
structure is expected around the burst, with highly ionized material
near the burst and progressively less ionized gas farther out, owing
to the geometrical dilution of the radiation field, the finite
propagation time of the ionization front,  and the opacity
of the circumburst medium. The ionization structure is also
time-dependent, as both the GRB flux and the integrated
opacity of the gas evolve. 

To model the behavior of the ionization front and structure with
time, we have performed a series of time-evolving photoionization
calculations using a code developed originally to measure the
evolution of ionized gas near Active Galactic Nuclei. The code solves
a set of linear differential equations that describe the evolution of
the abundance of the different ions $n_i$ as a function of the
variation of the incident flux with time, and as a function of the ionization and recombination processes (that also are time dependent). For a detailed discussion on time evolving photoionization processes we refer the reader to \citet[][]{nfp+99}and \citet[][]{kne+07}. 

The  code calculates the ionic abundances as a function of time for H
and He, as well as for heavier elements, namely C, N, O, Ne, Mg, Al,
Si, S, and Fe. It includes first order radiation transfer, as the
impinging radiation is attenuated by a series of optically thin layers
of material (with thickness $\sim 10^{16}$ cm for $n_{\rm H} = 1 \,
\rm cm^{-3}$), and is diluted by
geometry. The code does not include diffuse radiation (due to
recombinations) in the radiative transfer calculations (although they
are included in the time evolution of the ionic species). This is a
good approximation in the case of the material around GRB
afterglows because the expected densities in these regions
($\ll 10^6$ cm$^{-3}$) imply recombination times much longer than the
duration of the burst. For the conditions in this gas, the
recombination time ranges from tens to thousands of years  (to first
order, the recombination time is inversely proportional to the number
density of atoms in the gas). 

The code solves the time evolution in the first parcel of gas for the
whole light curve, then it uses the attenuated (time-dependent) flux
exiting this shell as the impinging flux for the next one. This
procedure is repeated until the ionization fraction is calculated at
all times up to a given distance. In our calculations we have produced
models up to 100 pc from the ionizing source. We only consider
distances larger than 1pc from the burst location. This has the
advantage of reducing considerably the computing time of the
models. At closer distances, the medium is completely stripped by 
the GRB radiation only a few seconds after the onset. A full description of
the code will be given in a forthcoming paper (Krongold et
al. 2013). Calculations using this code have previously been presented for
intrinsic absorption observed in the line of sight to  GRB
080330 \citep[][]{dfp+09}, GRB080319B  \citep[][]{dfp2+09}, and
GRB050922C \citep[][]{pwf+08}.  

As a simple test case, we present models for GRB
050730 and compare against an analagous calculation by \citet[][]{pdr+08}. The calculations were performed using the same parameters
reported by these authors,  namely, the luminosity of ionizing
photons, the light curve, and the spectral energy distribution (SED)
of the GRB afterglow, as well as the gas density (which is assumed to
be homogeneous with $n_H=10$ cm$^{-3}$). Figure \ref{fig:grb1}a (left panel)
presents our results.  The similarities between the ionization
structure predicted by our model (left panel in the Fig.) and the one
calculated by Prochaska et al. (Fig. 3 in their paper) at 1000 seconds
after the burst are reassuring. 
This figure also reveals the stratification of ionization states
within the surrounding medium.  In particular, it can be observed that the higher ionization charge states are formed closer to the source, up to a distance $\sim$ 15 pc away from the burst, where the ionization front is located.

In Figure~\ref{fig:grb1}b (right panel), we illustrate the time evolution for
several ions. In particular those high ionization ions that can be
more easily observed in the UV/Optical spectra of GRBs, as well as
OVII, which is a natural tracer of the highly ionized gas that may
provide a significant X-ray opacity. 
During the first few thousand seconds (observer frame time), gas up to 5-10 pc away
from the burst can reach a high level of ionization (showing 
states such as N V, O VI). At tens of kiloseconds, similar ionization
conditions would be driven to even larger distances (tens of pc). At these
times, the gas closer to the GRB will no longer be observed in these
ions because the gas will be driven to even higher ionization states
(e.g. N VI, O VII, O VIII) that are not observed in UV/optical
spectra but instead in X-rays. This plot exemplifies, in a simple way,
the expected time evolution of the ionization structure near the GRB.

We note in passing that observations of GRB
afterglows are typically carried out several hours (kiloseconds to tens of
kiloseconds) after the burst. Thus, if the hot gas is indeed
photoionized by the afterglow,  observations of highly ionized gas
do trace the close environment of the GRB, which may
include the circumstellar medium of the progenitor star and/or the
star forming region where it was embedded.      

\section{ A Fiducial GRB Model \label{sec:fiducial}}

Long-duration GRBs show a significant diversity in their spectral
properties and in the characteristics of intrinsic absorption
imprinted on their afterglows.  In a future work, we intend to explore
this diversity to infer variations in properties of the circumburst
medium.  In this manuscript, we focus on the specific challenge of the
set of GRBs which exhibit strong X-ray absorption with more modest
absorption in the optical/UV.  It is illustrative, of course, to consider
the problem quantitatively.  To that end, we consider a fiducial model
for a GRB event which is representative of many GRBs. Our fiducial
model (described below) is based on the average properties of the
sample of Swift GRBs presented by \citet{mzb+13} for which
data of the prompt and afterglow emission has been analyzed.  

\subsection{GRB Emission}

The GRB emission is characterized by a light curve which tracks the
time-evolution 
of the luminosity and a spectral energy distribution (SED) that may also vary with time.
The X-ray light curve can be described in general by two two different phases \citep[e.g.][]{wog+07}. Each phase is frequently
modeled and observed to follow a broken power-law $L \propto t^\alpha$.
The first phase (hereafter referred as the prompt phase) is strongly
related to the prompt $\gamma$-ray emission, and consists of a nearly
constant emission followed by a steep decay \citep{tgc+05,gtp+06}. In the second phase (hereafter the afterglow phase)
the light curve  flattens and then follows a ``normal decay" (the
break in this phase is generally interpreted as the shock break-out,
i.e.\ the 
time when the size of the relativistic beam matches the physical
extent of the jet). 

In the following, we model the prompt phase considering a constant
emission up to $t_p = 85$\,s after the burst followed by a steep decay
described by $\alpha_1 = -2.0$. We consider the transition between the
prompt and afterglow phases to occur at $t_b = 100$\,s. From this time
on, we model the afterglow light curve by a power-law with $\alpha_2 =
-1.0$. We do not model the break in the second phase from flat to
``normal" decay. This break usually involves a small change in slope
and takes place at later times, when the bulk of the photons has
already been emitted. Thus it has little effect in our results.  Our
choice of $\alpha_2$ and $t_b$ is based on the average values found by
Margutti et al. (2013). The rest of the parameters were constrained 
to match the average luminosities found for GRBs by these same
authors (see below).


Although the GRB emission is believed to be dominated by synchrotron
emission, its SED is both predicted and observed to be relatively
complex \citep{spn+98}.  One often models it as a series of
power-laws $L_\nu \propto \nu^{-\beta}$ across the electromagnetic
spectrum.  Our photoionization modeling, however, is only sensitive to
the SED at rest-frame energies from 1\,Ryd to 100\,keV, i.e.\ the UV to
X-ray passbands.  Therefore we model the SED as a single
power-law. We note however, that there can be strong spectral
evolution between the prompt and afterglow phases of the light curve
and that the early SED has not been studied systematically. However,
studies of individual cases show a much harder slope than that in the
afterglow \cite[e.g.][]{bph+06,spa+09}. In the following we assume $\beta = 0.0$ in the prompt
phase (before $t_b = 100$\,s) and $\beta = 1.0$ for the afterglow. 

Lastly, we have set the afterglow luminosity. We consider a total
X-ray energy $E_{X,iso} = 4.1 \times 10^{51}$\,erg integrated between
$0.3 - 10.0$\,keV and from $t=0$\,s to $t=10^4$\,s. The amount of
X-ray energy released in the prompt phase is $E_{X,pt} = 1.4 \times
10^{51}$\,erg and that in the afterglow is $E_{X,ag} = 2.7 \times
10^{51}$\,erg. These values are representative of the Swift GRB sample
by Margutti et al. (2013). With these considerations, the total amount
of ionizing photons emitted by this fiducial GRB  is  
$\phi_o = 2.1\times 10^{61}$ ph up to $t=10^4$\,s. 

\subsection{Absorption of the GRB Afterglow}

We compare the circumburst models ionized by the above afterglow to a
characteristic set of measurements for gas column densities from
analysis of X-ray and optical/UV afterglow spectroscopy.
Giving that the equivalent H column density obtained from the X-rays
and the column density of neutral material obtained from the optical
spectrum were derived assuming solar abundances (the neutral column
density assumes solar abundances as it was measured from metals, not
HI), we compare the results of our models  not to H, but rather to
O. 
Analysis of those GRBs exhibiting significant X-ray opacity 
typically yield estimates for the effective hydrogen column densities
of $N_{\rm H} \approx 10^{22} \cm{-2}$ (Watson et al. 2007; Schady et al. 2011; Campana et al. 2012).  These values were derived by
assuming solar abundances\footnote{The authors do not report the O
  abundance assumed in their models. However their models where
  carried out using Xspec. The standard abundances on Xspec can be
  varied according to different measurements, but all are consistent
  with the O abundance assumed here within a factor $\sim1.5$.} 
and a fully neutral medium.  The implied column densities of oxygen
therefore scale as $7 \times 10^{-4}$, giving $\nox \approx 10^{19}
\cm{-2}$.  We adopt this as our fiducial value for a typical GRB
showing X-ray absorption.

Regarding the optical/UV spectroscopy, the commonly observed
\ion{O}{1} transitions are always saturated and one can only estimate
$\nou$ by scaling measurements of non-refractory elements (e.g.\ S,
Si, Zn) assuming solar relative abundances.  Values from the
literature range from $\nou = 10^{16-18.5} \cm{-2}$.  As emphasized in
the Introduction, a significant
fraction of GRBs with X-ray absorption show $\nox \approx 10
\nou$ \citep{ssk+11}.
In the following, we adopt $\nou = 10^{18} \cm{-2}$ as our fiducial value.

Given that we will explore the conditions required to measure (much)
larger columns in the X-rays than in the optical, and these conditions
are likely due to ionization, we will evaluate our models considering
that the observed X-ray column density is produced mainly by the ionized gas, while the optical column is produced by the neutral one. As will be clear
from the models, this is indeed the case, and any contribution from the
neutral material to the X-ray absorption is negligible if larger
columns in this spectral region are required. Although we run models with many different conditions, we report here only the most relevant to explain the observations. 

\section{Results and Discussion \label{sec:disc}}

\subsection{Uniform Models of the Circum-Bust  Medium}
We begin by examining the time-dependent results from a series of idealized models with
constant gas density extending to a circumburst radius $\rcbm =
150$\,pc. Calculations at larger locations are not required given that the ionization front for all models takes place inside this distance.  
Within this model framework, we explore a range of gas density
($n=10$ and $n=10^3$ cm$^{-3}$) and 
metallicity ($Z=Z_\odot$ and $Z=0.01~Z_\odot$). 
Figure~\ref{fig:constdens} shows the radius, as a function of time,
to two transition regions driven by the afterglow for a series of
models.  These two transition regions are: 
(i) $\rhion$, the radius at which the gas
has a 50\%\ hydrogen neutral fraction; and 
(ii) $\roion$, the radius at which the gas has 90\%\ of its oxygen
completely stripped of electrons.   
These define the neutral region ($r > \rhion$) within the ambient medium
which dominates the UV opacity, and the start of the ionized region
($r=\roion$) where the X-ray bound-free and bound-bound opacity begins.
Both radii have a power-law dependence during the prompt phase emission ($t < t_b =100$\,s). At $t_b$ the afterglow emission begins and the radiation impinging on the gas becomes much softer ($\beta = 1.0$). This produces a rapid raise in $\rhion$ right after $t_b$, due to the larger number of UV photons. At later times ($t > 500$\,s)
both $\roion$ and $\rhion$ follow a power-law again, but with a different slope.

It is evident from the plot that there is relatively weak dependence
of the radii with metallicity as these are set by either the density
of the gas or the $r^{-2}$ dilution of the flux. 
One also notes that the evolution of $\roion$ 
is weakly dependent on the density.
This is because the gas is never optically thick to photons that
ionize O$^{+8}$ and $\roion$ is mainly set by $r^{-2}$ geometrical dilution 
of the afterglow.
Each $\roion$ curve reaches $\approx 3$\,pc at $t=10^4$s, indicating
the gas in the inner few pc cannot contribute to the observed X-ray
opacity.   
In contrast, $\rhion$ has a different evolution for the models considered.  
For models with higher
density, the ionization front lies closer to the source owing to the
greater opacity of the medium. Due to this larger opacity, in these
models  $\roion$ can even be larger than  $\rhion$ for the
few tens of seconds after the burst. This unusual ionization
structure, where \ion{O}{9}  can exist in the neutral H region is due
to the very
hard SED of the prompt emission phase. 


In Figure~\ref{fig:step}, we present the cumulative column
densities of ionized (solid) and neutral (dotted) 
oxygen at $t=1000$\,s for a series of models.  It is important to note
that the ionized columns include only the contribution from charge
states with at least one bound electron, e.g.\ 
\ion{O}{9} is not included.  Under the assumption
that the ionized gas must dominate the X-ray opacity, 
we require values of $\nox \approx 10^{19} \cm{-2}$. 
It is evident from Figure~\ref{fig:step} that
this requires a very high density ($n_{\rm H} > 10^2 \cm{-3}$), even
when one adopts a solar metallicity.  
If the large column densities observed in the
X-ray spectra of GRB afterglows  are indeed due to ionized material by
the burst, then the densities and/or metallicities of the
circumburst gas must be very large. {Alternatively, one could
  reproduce the observed $\nox$ 
values with neutral gas with much lower density extending to
circumburst distances $\rcbm
\gg 100$\,pc.  This would imply $\nou \approx \nox$ which is inconsistent with the observations.

An additional interesting point is obvious when comparing the model
with  $Z=0.01~Z_\odot$ and $n=10^3$ cm$^{-3}$ with that having
$Z=Z_\odot$ and $n=10$ cm$^{-3}$. Although the ionization front takes
place at different distances (Figure~\ref{fig:constdens}), the column
densities inferred for ionized O have similar values within a factor of a few. This is expected, as the total column
density of O depends almost linearly on both metallicity and
density. However, without an a-priori knowledge of where the
ionization front takes place (as is the case with spectroscopic
observations of GRBs), there is a degeneracy between the metallicity
and the density of the material. Because of this, we will discuss
models for solar metallicity in the rest of the paper, and will present those with
different metallicities when appropriate.


A uniform prediction of these constant density models is that the
column density of neutral material matches or even exceeds the ionized
phase within the first 100 pc from the burst.  This follows from the fact that $\rhion \lesssim 20$\,pc 
(Figure~\ref{fig:constdens}) such that the medium is dominated by 
 neutral gas along any given sightline.
Integrating to $r=100$\,pc, all the models predict larger columns of
neutral than ionized O (by at least a factor of a few). As such, one
predicts column densities probed by X-rays that exceed the optical/UV by only
as much as 50\% ($\nox < 1.5 \nou$). 
We note that it is likely that the ISM of the host galaxy extends further beyond the first 100 pc from the burst, making the X-ray and optical/UV columns even more similar.  This violates the constraints of our fiducial afterglow model where $\nox \sim 10\nou$. 
Thus, unless the circumburst medium has size $\rcbm \ll 100$\,pc, 
this model contradicts what is frequently observed.
In this case, the bulk of the opacity (in both wavelength ranges) would be
dominated by neutral material. 
%
We conclude that if the large columns found in the X-rays
are indeed due to material ionized by the GRB afterglow, then the
density and metallicity should be large, and the distribution of
material cannot be homogeneous. 

Given that the the progenitors of long GRBs might be massive stars
with stellar winds \citep{Woo93}, possibly embedded in intense
star-forming regions, we must consider the possibility that the 
material within a few
tens of pc from the progenitor was ionized prior to the onset
of the burst.  Indeed, simple treatments of likely progenitors for
GRBs predict such `pre-ionization' \citep{pl02,wph+08}.
Then, it is important to test how these conditions may affect the
results presented above. We have produced a series of 
models varying $\rpre$, the radius to which the gas has been
pre-ionized prior to the GRB event. We assume an initial ionization
distribution consistent with gas ionized by radiation following the
spectral energy distribution of a massive star ($M \sim 30 M_\odot$) 
giving a H neutral fraction of $f(HI)\sim 10^{-5}$.  
For $\rcbm \approx 100$\,pc, values of $\rpre < 50$\,pc yield very
similar results to those models without pre-ionization, as can be seen in Figure~\ref{fig:step} (magenta line). This is because, although the ionization front and the ionized columns of material are larger due to the effects of the pre-ionization, at distances  $r\sim 2\rpre$ from the burst  the neutral column becomes comparable to that of ionized material. We note that  $\rpre$ might actually be much larger, satisfying
the condition $\rpre \approx \rcbm =100$\,pc. In this case, the
contribution to the total column density from the neutral material up
to $\rcbm$ would be negligible (given that this gas was already
ionized by the progenitor). However, as before, at a distance $\sim
2\rpre$ the neutral material from the ISM of the host
galaxy would contribute as much as the ionized material to the total
column density, so that again $\nox \sim \nou$. Then, in order
to have $\nox \sim  10\nou$, the total optical path length from the
burst location to the ending radius of the galaxy would have to be
$\sim \rcbm$. While this might be indeed the case for a few GRBs, we
consider this possibility unlikely, because it requires that most GRBs
are located near the edge of their host galaxies contrary to
observations \citep{bkd02,fls+06}.


  \subsection{Circum-Bust Medium Models Including Gradients of Density \label{grad}}

Homogeneous models predict similar column densities in the X-ray and
the optical/UV domains, with the bulk of the opacity due to neutral
material at distances larger than a few tens of pc from the GRB.
Furthermore, high column densities of ionized material require
a circumburst medium with large density (or metallicity).   
Motivated by these results, we have explored simple models with
gradients of density as a possible explanation for the larger X-ray
column densities. The models presented here assume larger densities
close to the burst, decreasing outwards. We explore models with
a step in the density: $n=n_1$ at $r\leq \rstep$ and  $n=n_2$ at $r>
\rstep$,  with $n_1>n_2$. In all models we assume $n_2 = 1 \cm{-3}$.
Although this model is overly simplistic, models with radial gradients in density give similar results.

Figure \ref{fig:step1} presents our results. The most important difference with the homogeneous cases is that, in general, in the step function model the neutral column density can be, by design, much lower than that for the constant density models. This is clearly observed when comparing the models with $n_1 = 10^3 \cm{-3}$ (blue and cyan lines in Fig.  \ref{fig:step1} and blue line in  Fig. \ref{fig:step}).

Indeed,
some step function models do predict  $\nox \sim  10\nou$. For
instance, a model with $n_1 = 10^3 \cm{-3}$ and  $\rstep = 5$\,pc
produces X-ray and UV column densities similar to the fiducial
afterglow values (blue line in Fig.  \ref{fig:step1}). However, if the
same model is considered, but with  $\rstep = 10$\,pc (cyan line),
then $\nox \sim \nou$ and this condition is no longer satisfied. This
is because in the second model $\rstep > \rhion$, and the dense region
can contribute significantly to the neutral absorption. In general,
these models are only capable of producing larger columns in the
X-rays than in the UV if  $\rstep \sim \rhion$. Since there is no
physical reason to assume a relation between the size of the
ionization radius (which depends on the GRB properties) and the size
of the dense region, we conclude that these models cannot naturally
explain the ensemble of observations. 
This motivates a further addition to the model.

\subsection{Models Including Gradients of Density and Pre-Ionization in the Circum-Bust Medium}

Finally, we explore models with gradients of density, considering that the medium surrounding the GRB was already ionized before the burst explosion. As discussed before, there are strong arguments to assume a pre-ionized medium \citep{pl02,wph+08}. We explore the same density distribution as in \S \ref{grad} ($n=n_1$ at $r\leq \rstep$ and  $n=n_2$ at $r>\rstep$,  with $n_1>n_2$ and $n_2 = 1 \cm{-3}$). 
To avoid cases where  $\rstep > \rhion$, which produce $\nox \sim
\nou$  (\S \ref{grad}), we consider only models where the region of
pre-ionized gas contains the dense cloud of material (that is $\rpre >
\rstep$). As before, we assume a H neutral fraction $f(HI)\sim 10^{-5}$.
Figure \ref{fig:step2} presents our results. There are  two notable benefits in these models.

\begin{enumerate}
\item Given the pre-ionization condition, the column density of the ionized material is
dominated by gas at $r < \rstep$.  Therefore, the value of $\nox$ measured is
determined by the density and size of the denser region. 

\item The neutral column density is produced, by
design, outside the denser region. Therefore,  $\nou$ is systematically lower than the column of ionized gas, i.e.\ $\nox \gg \nou$. 
\end{enumerate}

We find that a model with $n_1 = 10^3 \cm{-3}$ and
$\rstep = 10$\,pc present X-ray and UV column densities similar to the
fiducial afterglow values. The cyan line in Figure \ref{fig:step2} presents this model. This is not a unique
solution, of course; many solutions are possible, with lower $n_1$
requiring larger $\rstep$. However, the solutions must satisfy two
important conditions: (1) $\rstep \lesssim \rpre   < \rhion$, so that there is
no contribution from the dense region to the neutral column density,
and (2)  we require $\rstep > \roion$ so that the GRB does not
entirely ionize the dense medium that provides the X-ray
opacity. 

Within these restrictions, the values that can
reproduce our observations are constrained to be in the ranges 30\,pc
$> \rstep > 5$\,pc, and   $10^4  \cm{-3} > n_1 > 5\times
10^2\cm{-3}$. Different solutions produce X-ray and
optical/UV column densities much different than the ratio of 10 from our
fiducial value. For instance, a model with $\rstep = 20$\,pc and $n_1
= 10^3 \cm{-3}$ (blue line in Fig. \ref{fig:step2})
would easily produce two orders of magnitude larger column in the
X-rays than in the optical/UV, as found in several objects \citep[][]{ctu+10}.
These restrictions also impose a minimum pre-ionization radius around the GRB progenitor (condition (1) above), regions with  $\rpre \sim 20-30$\,pc are enough to explain the observations (see Figure \ref{fig:step2}). We consider these values to be very conservative, since detailed modeling shows that the progenitor can  pre-ionize the surrounding material at distances $\gtrsim 100$\,pc  \citep[][]{wph+08}. 
We note that our results do not place much restriction on the value of $n_2$
(the only condition being to keep small the neutral column density). Assuming again that the neutral gas in the ISM extends for at least $\sim 100$\,pc beyond the circumburst region implies that $n_2 \lesssim 10  \cm{-3}$.

Given that step function models are very simplistic,
we have also computed models assuming a constant density up to a
distance $\rstep$ from the burst, and a density law parameterized by a
power-law($n=n_o(r/\rstep)^{-\alpha}$) at larger distances. We find 
qualitatively similar results to the step models provided $\alpha
\gtrsim 2$.  These models also yield $\nox > \nou$ for $\rstep <
30$\,pc. 


We conclude that the different column densities found in the X-ray and
optical/UV regimes of GRB afterglow spectra can be well explained by
the presence of ionized material within a few tens of pc of the burst location. This requires that the GRBs are produced in dense clouds of material within the host galaxy, with column densities characteristic of molecular clouds, and that this material is ionized prior to the explosion \citep[][]{wph+08}. Our results even permit
ionized versus neutral column density ratios up to 2-3 orders of
magnitude or more, as reported by \citet[][]{ctu+10}. We note that some evidence of evolution with cosmic time in the X-ray absorbing column has been reported in previous works \citep{ctu+10,bdd+11}.  We plan to test different scenarios to explain this evolution in a forthcoming paper\footnote{While our manuscript was in revision, we became aware of a recent work by \citet{wza+13} suggesting that absorption by He II may be producing most of the X-ray column density, and may be responsible for its evolution. This idea requires a SED with a very hard X-ray to UV photon index  ($\Gamma \sim -1$), much harder than what is usually  observed in GRB afterglows. A full time-evolving photoionization model testing this idea is warranted.}.

\subsection{Additional Considerations on Mass and Metallicty}

An independent constraint on the metallicity of the medium surrounding
the GRB may come from stellar evolutionary theory. In this theory,
late-type massive stars, such as Wolf-Rayet stars, are considered the
most likely progenitor candidates for GRBs \citep{wh06}. 
Furthermore, low
metallicities ($Z\lesssim 0.1- 0.3 Z_\odot$) are preferred by the
models to avoid the spindown of the stellar core \cite{irt04,yl05,wh06}. 
We note however, that it is possible that the circumburst medium of the GRB has been 
contaminated with metals by the progenitor (at least to some
extent). Thus, an extremely low metallicity is not strictly required
to reconcile stellar evolutionary results with those presented
here. However, somewhat low metallicities may still be expected in the
circumburst region, and thus they are definitely desirable in the
models. We note that it is possible to get large columns of ionized
material with arbitrarily low metallicities, provided that the density
of the clouds where the progenitors are embedded is large enough. 

Nevertheless, large densities may
imply unphysically large masses for the progenitor's cloud. For the more massive model presented in  Figure \ref{fig:step2} (blue line), the implied mass
for $\rstep = 20$pc is $\approx 10^6 M_\odot$ . 
This mass is characteristic of molecular clouds within the Galaxy
\citep{m11}.  Even a circumburst medium with $Z=0.1 Z_\odot$ may
match our fiducial value for $\nox$ (red line in
Fig. \ref{fig:step}).  Only if $Z \ll 0.1 Z_\odot$ \citep[e.g.\
GRB~050730][]{cpb+05}
would one require significantly larger density 
($n_1 > 10^4 \cm{-3}$, magenta line) and a correspondingly larger mass of the
circumburst medium.  To the best of the authors knowledge, no GRB has been detected with such a low metallicity medium which also shows significant X-ray
opacity.

We conclude that the cloud's radius, density, and mass required to
explain the differences between the X-ray and optical/UV column
densities may be large (particularly if low metallicities are also
required), but are fully comparable with those found in giant molecular
clouds. Thus, our results support a scenario where GRB's explode in
giant molecular clouds within their host galaxies \citep[][]{crc+06,pdr+08,ctu+10}. A statistical
study with observations of a large number of GRBs, spanning a wide
range of luminosities, as well as X-ray and optical column densities,
is required to further explore the properties of these clouds.     


\subsection{Ionic Column Densities and Time Variability}

Two key ingredients in testing any model of the circumburst medium around
GRBs is to study (1) the column densities of individual charge states, and (2) the possible variability (or lack of it) in the overall absorption properties of the material. 
 In the following, we only discuss step function models including a pre-ionized medium, as these are the only ones consistent with the different X-ray an optical/UV column densities. We note that models with gradients of density are more likely to produce variability than models with
constant density,  because a charge state that  was produced
within the dense medium surrounding the GRB at short times after the
burst, might be produced outside this region at later times. This
would result in strong changes in column density for that particular
ion. 

To exemplify this,  in Figures \ref{fig:si_4}, \ref{fig:c_4}, and \ref{fig:n_5} 
we present the time evolution of the fraction of Si IV, C IV, and N V for two different models with $n_1 = 10^3 \cm{-3}$ and $Z=0.1~Z_\odot$. In the left upper panel  of the
plots (where a model with $\rstep=15$\,pc is presented) one observes that 100\,s after the burst, the vast majority of these ions are produced in the dense medium. However, when the observed time is $\sim1000$\,s, they are fully produced at $>
15$\,pc, in the low density regions. Between these two times, the column density of these charge states decreases rapidly, as the peak in their fractional distributions moves out from
the dense region into the diffuse medium (see solid lines in the bottom panel of Figures \ref{fig:si_4},  \ref{fig:c_4}, and \ref{fig:n_5}). 

An important additional constraint on the structure of the dense
clouds of gas is set by the measured column densities of \ion{N}{5}, which
traces more highly ionized gas. The UV transitions from this ion have
thus far been observed to be unsaturated, with column densities  
$N(N^{+4}) \sim 10^{14}$\,cm$^{-3}$ (e.g. Prochaska et al. 2008, Fox et al. 2008). 
This imply that N\,V must be produced outside of the dense regions
during these observations. In the case of our fiducial model, this
means that the maximum allowed dense region size is
$\rstep=15-20$\,pc, 
because for larger sizes N\,V would be produced inside $\rstep$ even
thousands of seconds after the burst, and the predicted column
densities would be $\sim 100-1000$ times larger (see Fig.
\ref{fig:n_5}). However, we note that the GRBs where N\,V and other
high ionization UV charge states can be observed are distant
($z\gtrsim 2$), and $>10$ times brighter than our fiducial GRB
luminosity (which is based on the average luminosity distribution by
Margutti et al. (2013) and includes objects at all
redshifts). Indeed, a model with 5 times more ionizing photons than
our fiducial GRB produces N\,V at distances larger than 40 pc, only
1000\,s after the burst (see upper right panel and dotted line in the
bottom panel of Fig. \ref{fig:n_5}). This makes possible much larger
sizes for the dense region. In these bright GRBs, absorption from
different charge states can be produced at large distances from the
burst (hundreds of pc or more depending on the GRB luminosity),
consistent with the observations (e.g. Fox et al. 2008).  This is
because of the much smaller opacity in a pre-ionized medium, with
respect to a neutral one. Thus, pre-ionization is not only required to
explain the large X-ray to optical/UV column density ratios, but also
to explain absorption by photoionized elements at large distances from
the GRB.  

We remark that our models predict no variability in the total column densities of optical/UV observed charge states at times $> 500s$, when the earliest spectroscopic observations in these wavebands are possible \citep[e.g.][]{flv+08, dfp+09}. Our models, however, are not in contradiction with possible variations in individual charge states, as these can be produced by additional inhomogeneities in the diffuse medium, beyond $\rstep$. These models are also consistent with the observations of variability in excited lines \citep[e.g.][]{vls+07,dfp+09}, as these lines arise from material at large distances from the GRB region, in the ISM of the host galaxy. The spectral variations are produced by UV pumping from the GRB afterglow radiation into the excited levels of a single ion  \citep[e.g.][]{pcb06,dfp+09} and not by variations in the total column density of that particular ion.

Finally, it is interesting to study how the total column density would vary with time. In Figure \ref{fig:ion_var} we present the time evolution of the X-ray column
density for the same models presented in Figure
\ref{fig:step2}.  The plot shows the expected variability for ionized O
(excluding again \ion{O}{9} that does not contribute to the
opacity). Only mild variations in the column density of ionized O should result
in small variations in the measured column density in multi-time X-ray spectra of the afterglows. As it can be observed, the different step models predict variations by a factor $\lesssim2$, according to their  $\rstep$. Variations by these factors cannot be measured in actual X-ray data (given the limited signal to noise ratio in the data and the uncertainties in the modeling). Nevertheless, a trend is expected with smaller columns found at larger times. Larger column density variations may be expected for GRBs embedded in smaller radii clouds (with $\rstep\sim\roion$=5\,pc), as most of the gas cloud would  be fully ionized at later times. However, this would also imply much smaller X-ray to Optical/UV column ratios at later times.

\section{Summary and Future Work\label{sec:conc}}

In this work we present time-evolving photoionization calculations of the environments of GRBs. 
Our models indicate that the discrepancies found in the column density measurements carried out in the X-ray and optical/UV domains can be easily explained in terms of 
strong density gradients between the burst region, where the ionized
X-ray absorption arises, and the extended (neutral) ISM of the host
galaxy (imprinting absorption features in the optical/UV
band). However, this requires that the dense medium surrounding the
GRB is pre-ionized, as suggested by previous works. 

Our results support a scenario in which the bursts are produced in dense condensations of material within their hosts. The masses, radii, and densities of these clouds are consistent with those found in giant molecular clouds, further implying that GRB are indeed produced within regions of intense star formation  in galaxies. A case study for a fiducial GRB is too limited to further provide insights on the overall properties of such clouds (sizes, densities, masses, and metallicities distributions). However, here we show that time resolved spectroscopy of X-ray and Optical/UV data have strong diagnostic capabilities when compared to simple models.  Producing detailed photoionization models --with physical parameters able to match the measured X-ray and optical column densities, for a large sample of bursts-- is a promising way to shed light on the physical properties of the star formation regions where the GRB progenitors are born and evolve, and also on the ISM of their hosts.

\acknowledgements 

We thank the anonymous referee for thoughtful comments that improved our paper and
Enrico Ramirez-Ruiz for insightful discussions.
JXP acknowledges support from 
NASA/Swift grants
NNX07AE94G and NNX12AD74G. YK acknowledges support from CONACyT 168519 grant  and UNAM-DGAPA PAPIIT IN103712 grant. 
YK and JXP acknowledge support from a UC Mexus grant, FA 10-61.


\bibliographystyle{apj}
\bibliography{grb_references}


\clearpage

\begin{figure}
\plottwo{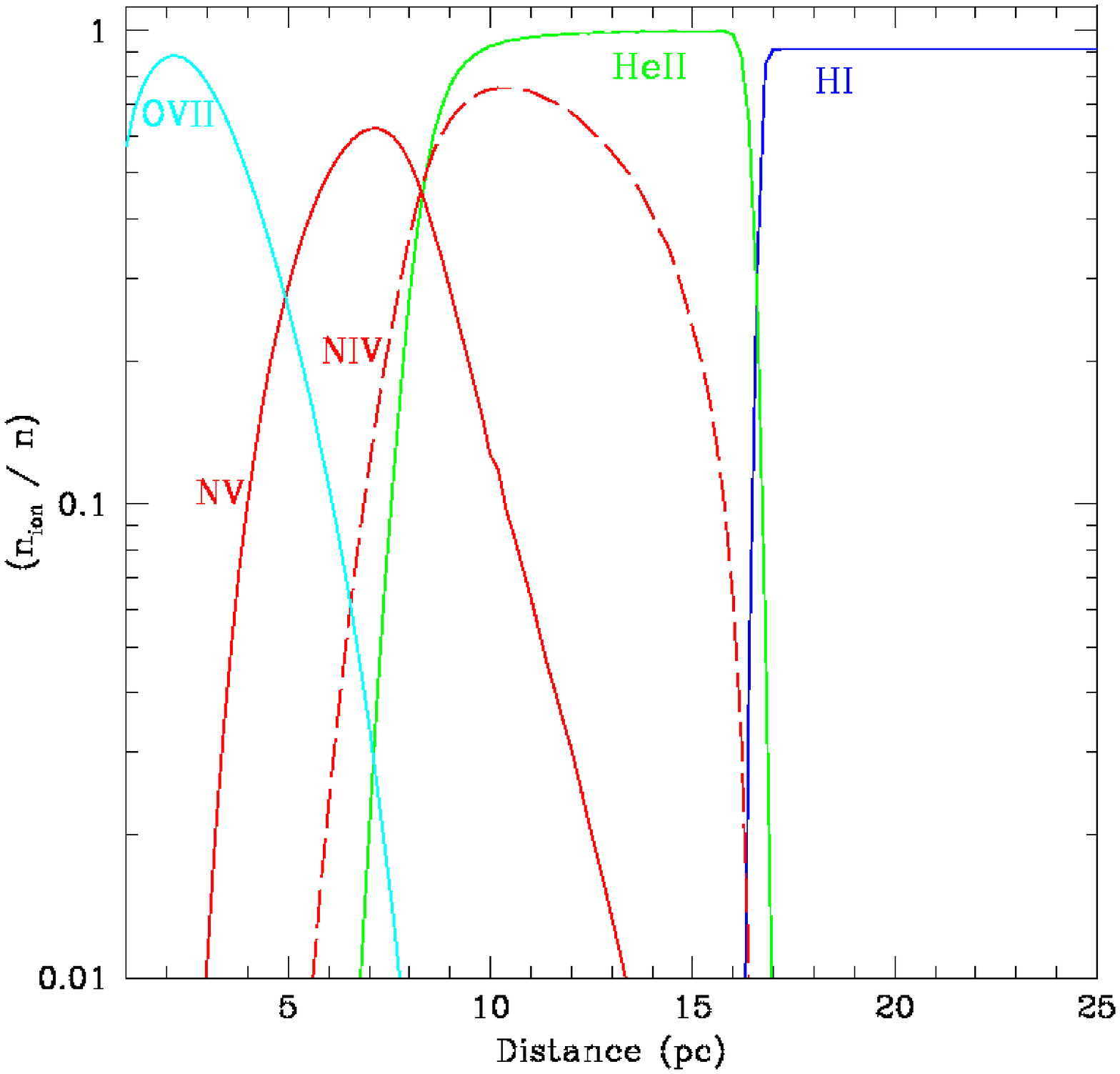}{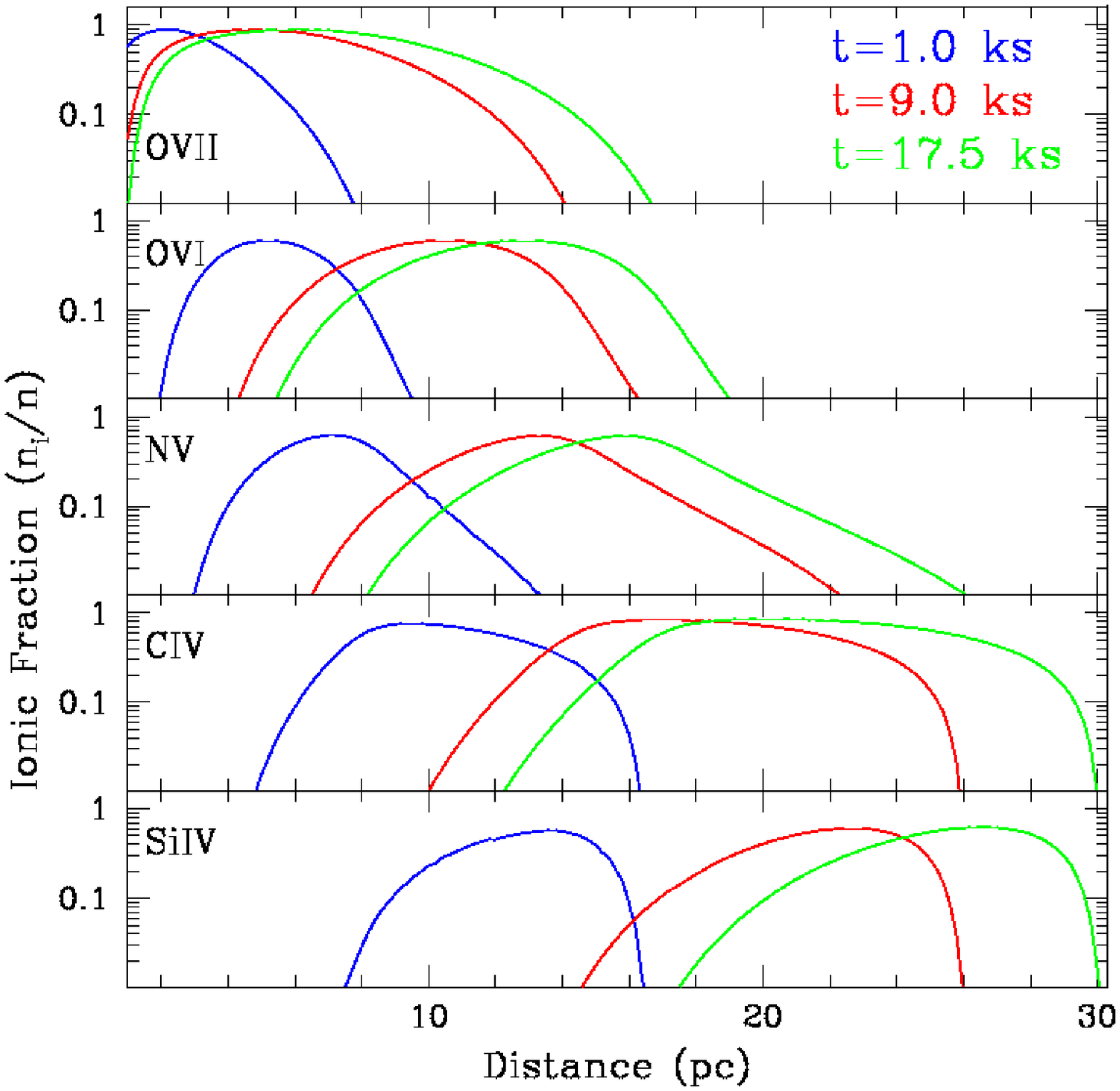} 
\caption[]{Time dependent photoionization model of the ISM
  surrounding GRB050730. The model was produced assuming the same
  conditions used by (Prochaska et al. 2008), including an homogeneous
  medium with number density n(H)=$10$ cm$^{-3}$.  This comparison
  provides a test of the two codes and we find excellent agreement in
  the results.
  (Left): Ionization
  structure at observed time t=1 ks. (Right): Ionization fractions for
  high ionization charge states as a function of distance at three
  different times, showing the time evolution of the ionization
  structure. \label{fig:grb1} } 
\end{figure}



\begin{figure}
\plotone{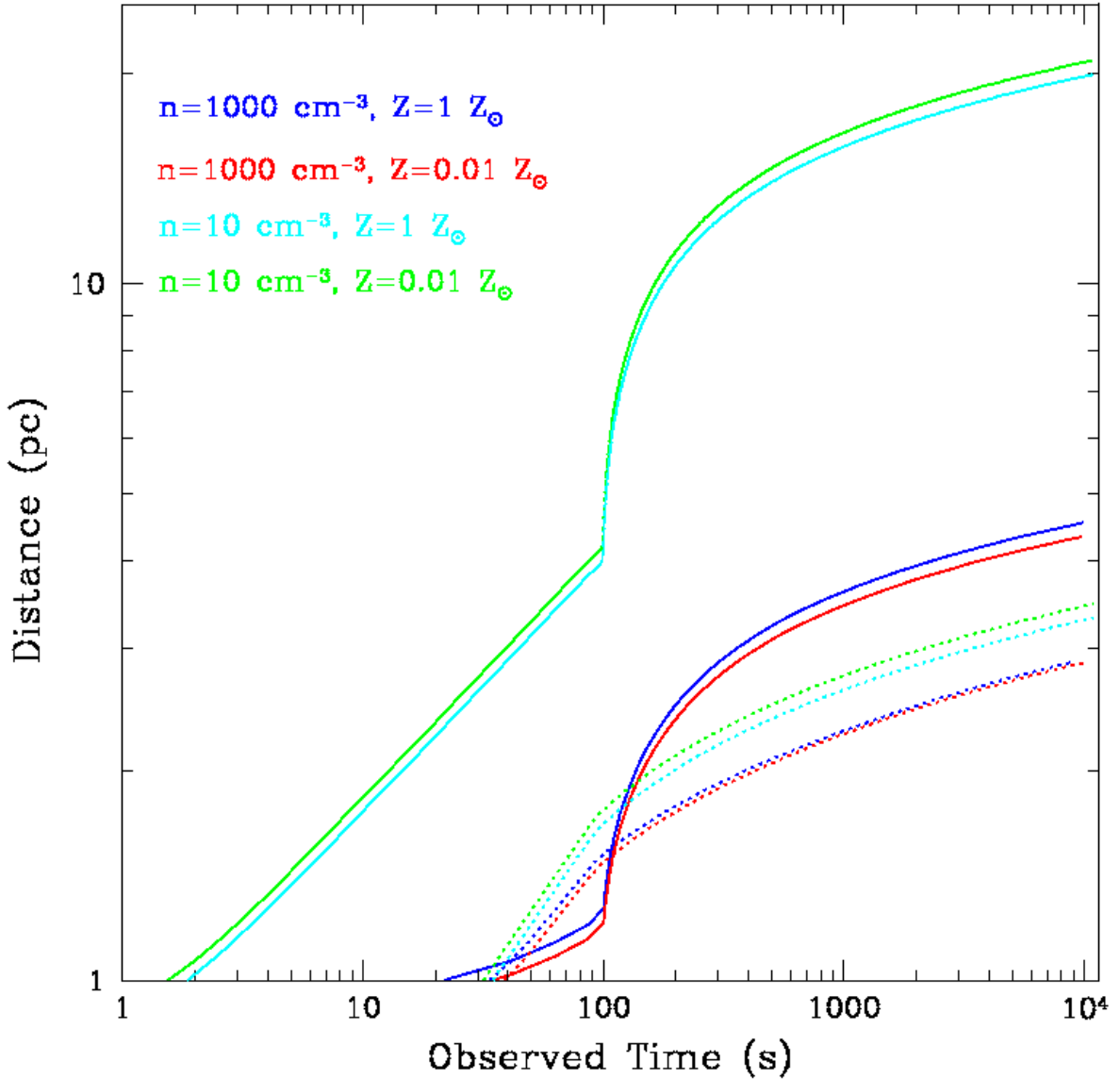}
\caption[]{Time-dependent evolution of the ionization fronts of two transition regions: 
(i) solid curves trace the radius $\rhion$ at which the gas
has a 50\%\ hydrogen neutral fraction; and (ii)
dotted curves trace $\roion$, the radius at which the gas has 90\%\ of its oxygen
completely stripped of electrons. gas. The four models presented
include two densities with two  
different metallicities each.   We note very weak metallicity
depedence on the results and find strong density variations only for
$\rhion$.   In general $\rhion$ exceeds $\roion$ as expected, but the
prompt emission radiation field is sufficiently hard that does predict
$\roion > \rhion$ for short times in some models.
\label{fig:constdens} } 
\end{figure}


\clearpage

\begin{figure}
\plotone{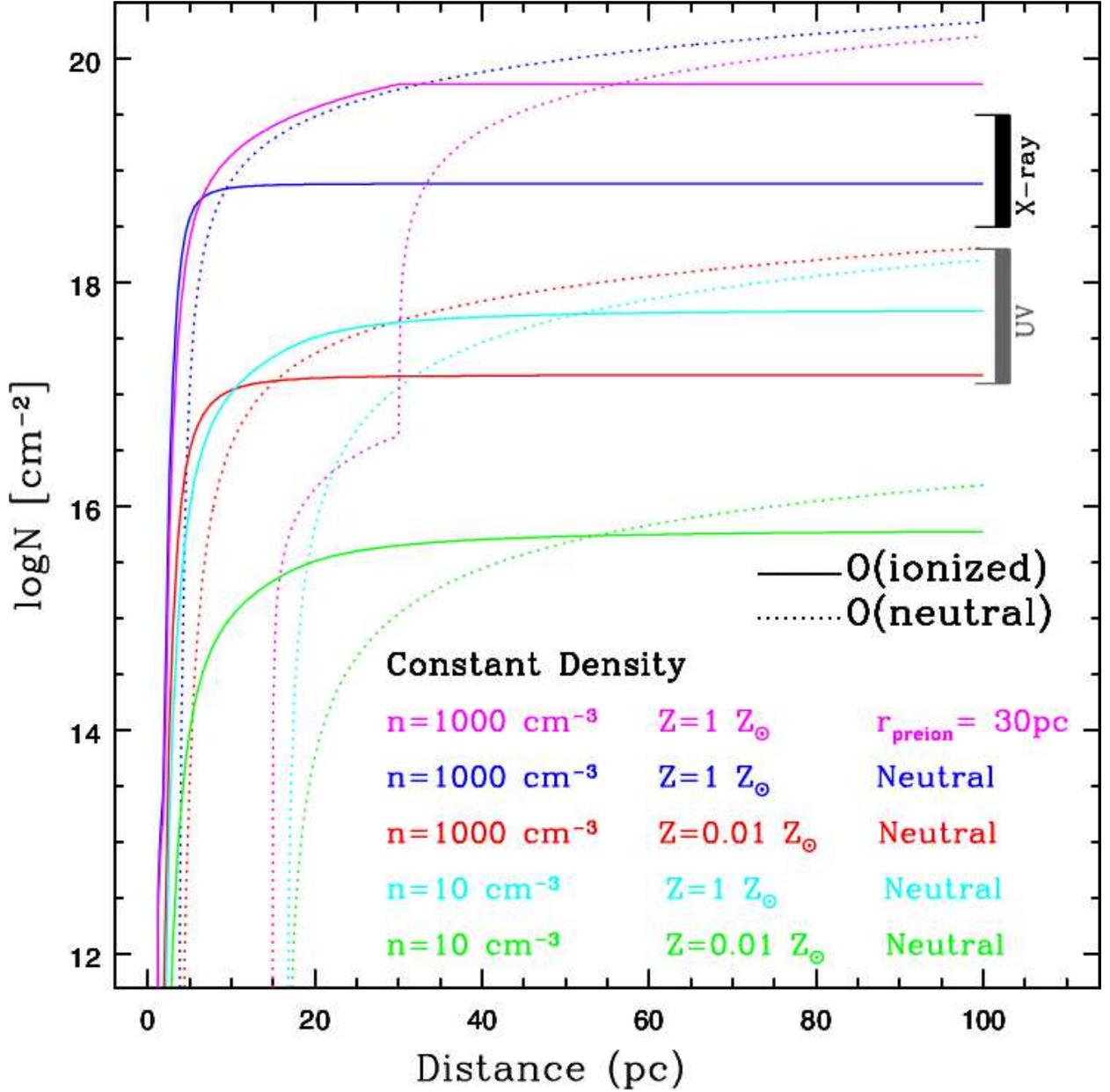}
\caption[]{ Predicted column densities from our time-dependent
  photoionization calculations of the circumbust medium at $t=1000$s.  These
  results are for a series of models assuming an homogenous medium,
  i.e.\ constant density and metallicity.  Such models predict similar X-ray and 
  optical/UV column densities.  
  Generally, we find that the column densities of the gas which dominate
  the UV opacity are comparable to or even exceed those which are
  expected to produce the majority of X-ray opacity.  Therefore, we
  consider these constant density models to be unviable.
\label{fig:step}} 
\end{figure}


\clearpage

\begin{figure}
\plotone{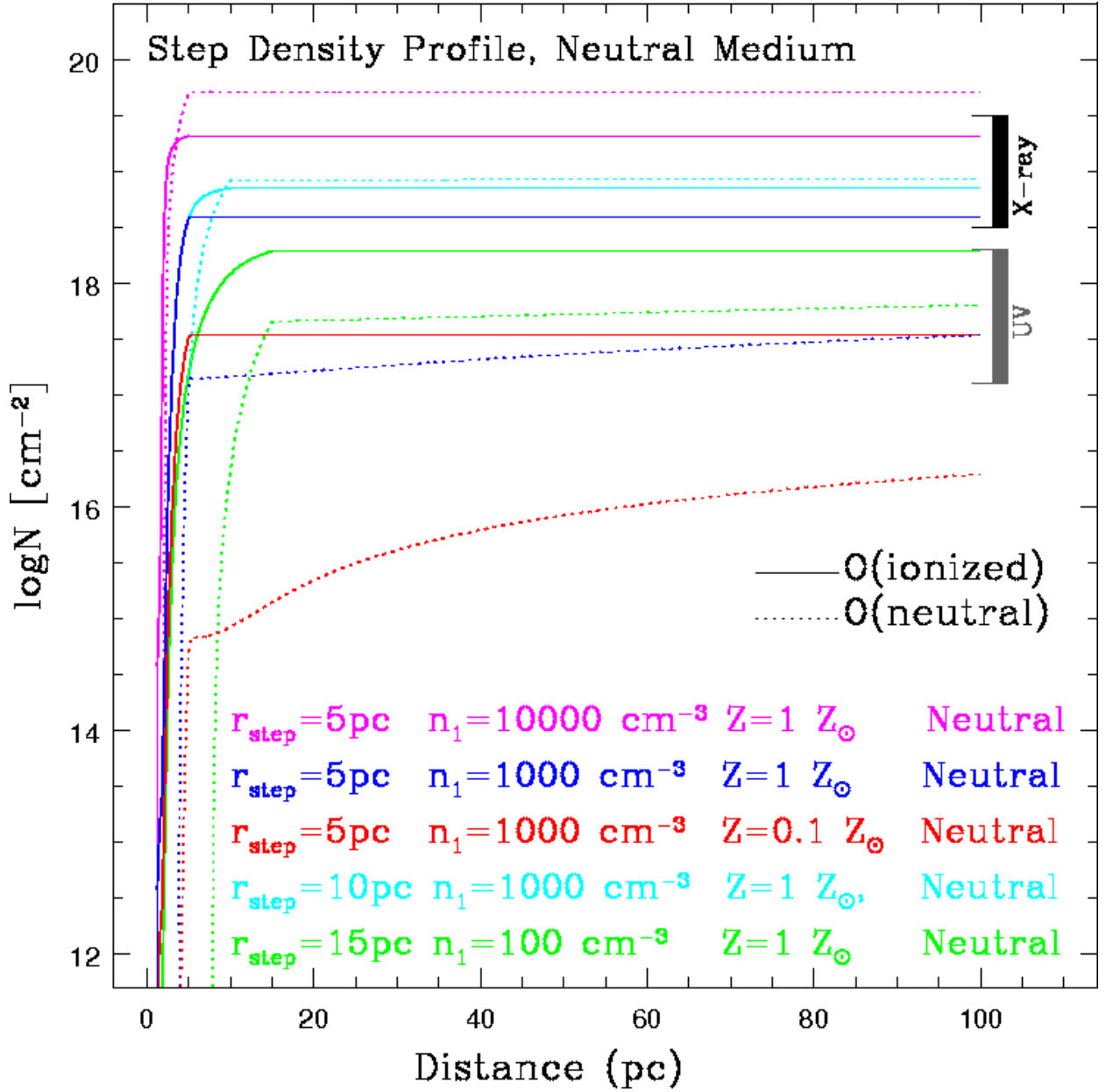}
\caption[]{ Predicted column densities from our time-dependent
  photoionization calculations of the circumbust medium at $t=1000$s.
  The results are for models consisting of strong
  gradients of density between the regions close to the burst location
  (n$_1$) and the ISM (n$_2=1$\,cm$^{-3}$). A neutral medium prior to
  the burst is considered. Only models where the size of the dense
  regions is comparable to the ionization front can produce large
  X-ray to optical/UV column density ratios (e.g.\ $r_{\rm step} =
  5$\,pc, blue-colored curve).
\label{fig:step1}} 
\end{figure}


\clearpage

\begin{figure}
\plotone{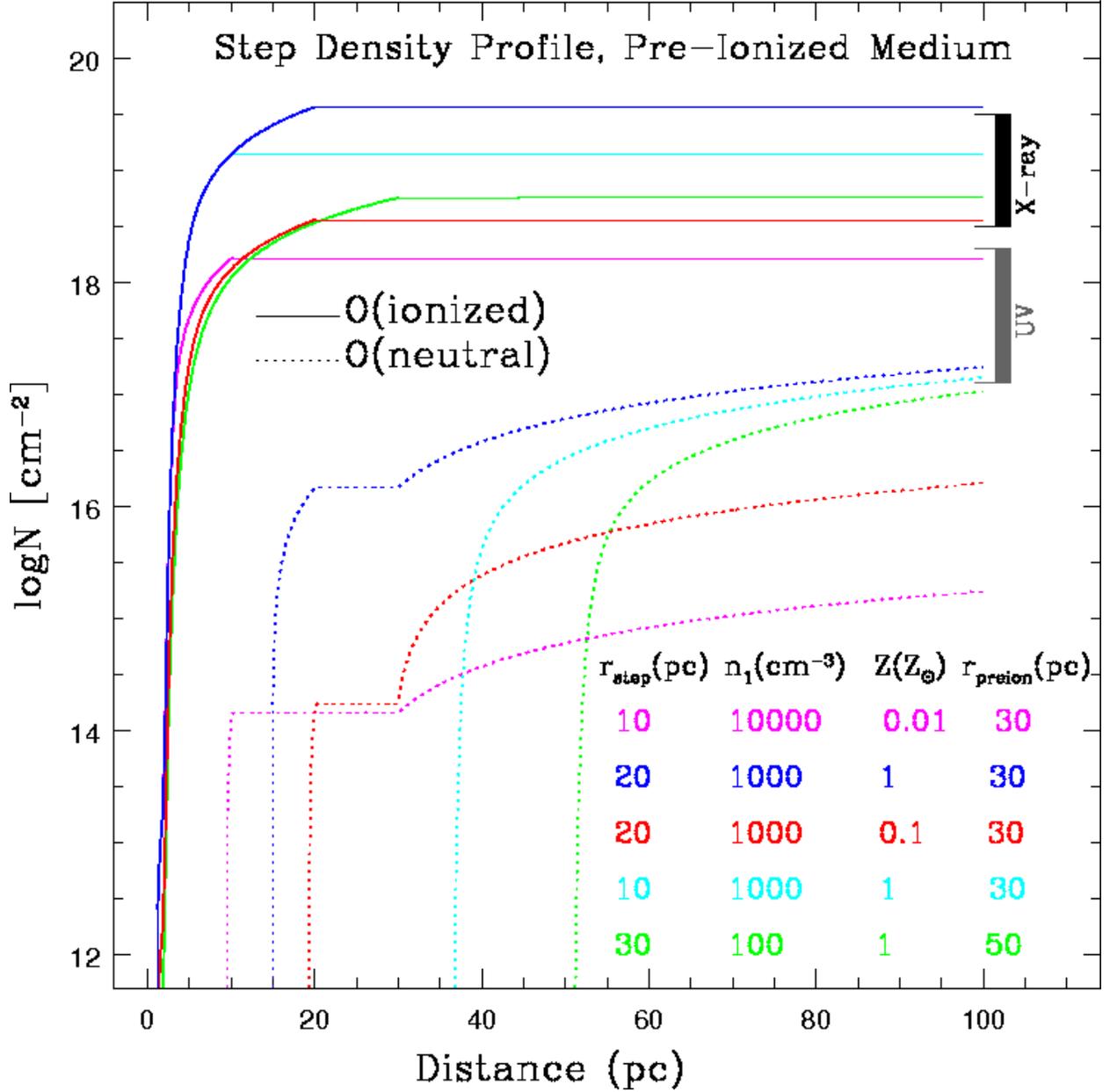}
\caption[]{ Predicted column densities by our time-dependent
  photoionization calculations of the circumbust medium  at $t=1000$s.  These cruves
  show the results for models consisting of strong
  gradients of density between the regions close to the burst location
  (n$_1$) and the ISM (n$_2=1$\,cm$^{-3}$) and also impose 
  a pre-ionized medium where the gas within $\rpre$ has been ionized
  to have a H neutral fraction $f({\rm HI}) = 10^{-5}$. 
  All these models are able to explain the large
  X-ray columns, as well as the `excess" with respect to the
  optical/UV values, as required by the observations.    This is
  provided that $\rpre = 30$\,pc.
\label{fig:step2}} 
\end{figure}


\clearpage

\begin{figure}
\plotone{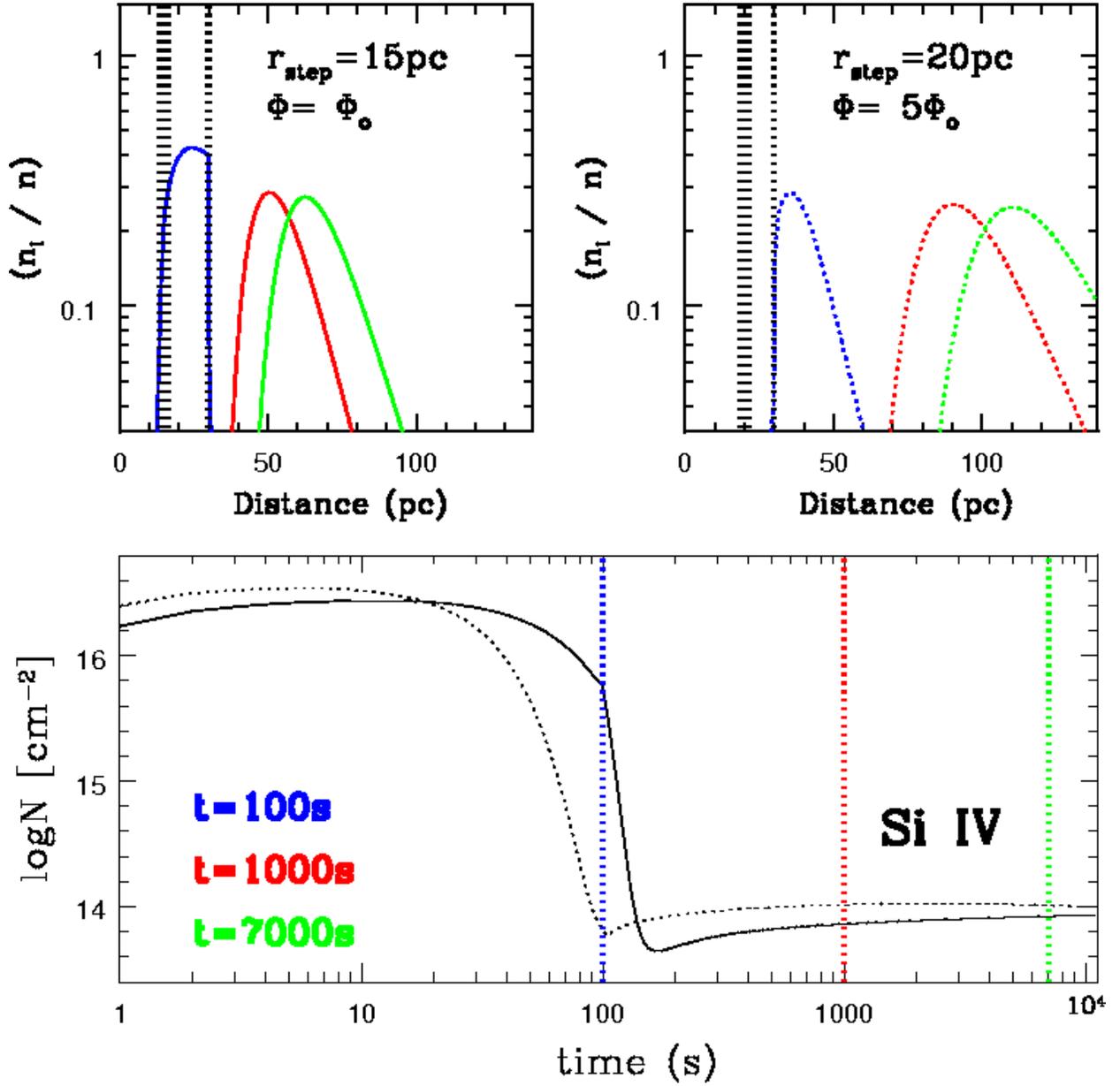} 
\caption[]{Time evolution of Si IV. The upper panels show the spacial
  distribution of the fraction of Si IV as a function of distance to
  the location of the burst, at three different times after the
  event. Two models are presented consisting on  a step 
function with n=$10^3$ cm${-3}$, and  Z=0.1\,Z$\odot$. The models
further assume a pre-ionized medium with $\rpre=30$\,pc. The left
panel shows a model with $\rstep=15$\,pc, and $2\times 10^{61}$
ionizing photons (our fiducial GRB). The right panel considers a case
with $\rstep=20$\,pc, and five times more ionizing photons. 
The lower panel presents the column densities of Si IV as a function
of  time for the two models. The vertical color lines mark the
ionization distributions at the times presented in the upper panels.   
\label{fig:si_4} } 
\end{figure} 



\clearpage

\begin{figure}
\plotone{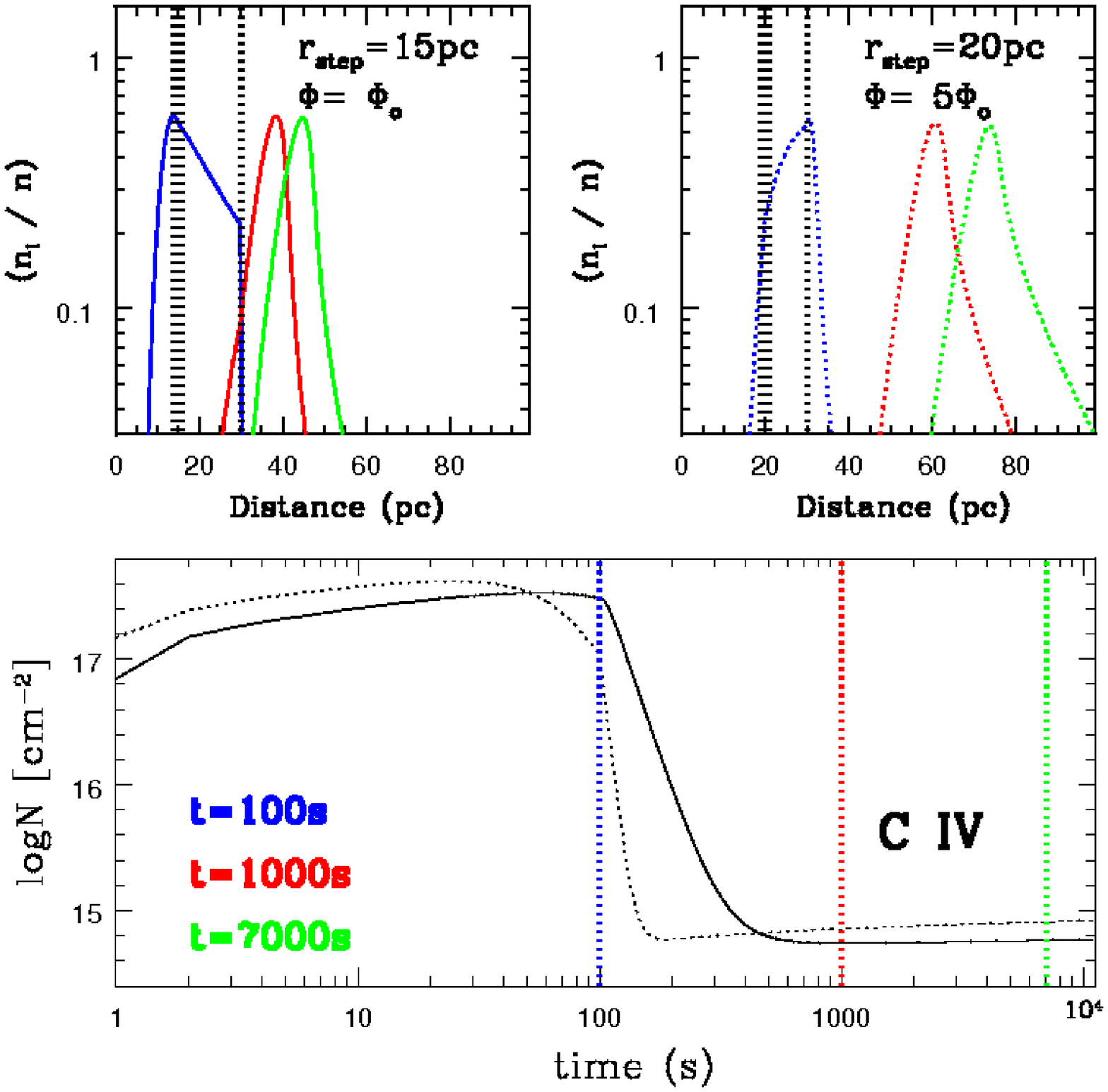} 
\caption[]{Time evolution of the fractional abundance and column
  density of C IV. Models and labels as in
  Fig. \ref{fig:si_4}.  
\label{fig:c_4} }  
\end{figure}



\clearpage

\begin{figure}
\plotone{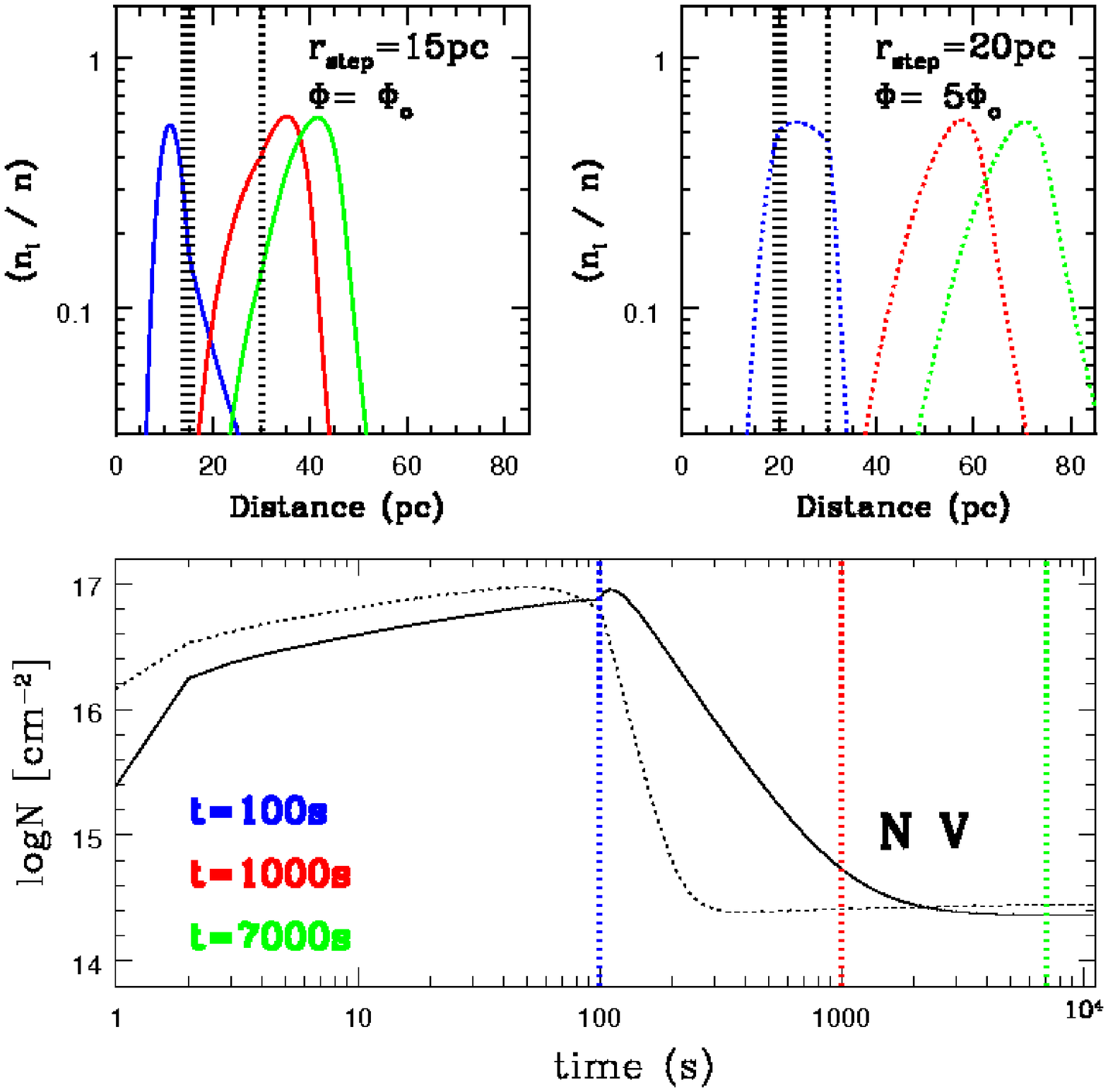} 
\caption[]{Time evolution of the fractional abundance and column
  density of N V. Models and labels as in
  Fig. \ref{fig:si_4}  
\label{fig:n_5} }  
\end{figure}



\clearpage
\begin{figure}
\plotone{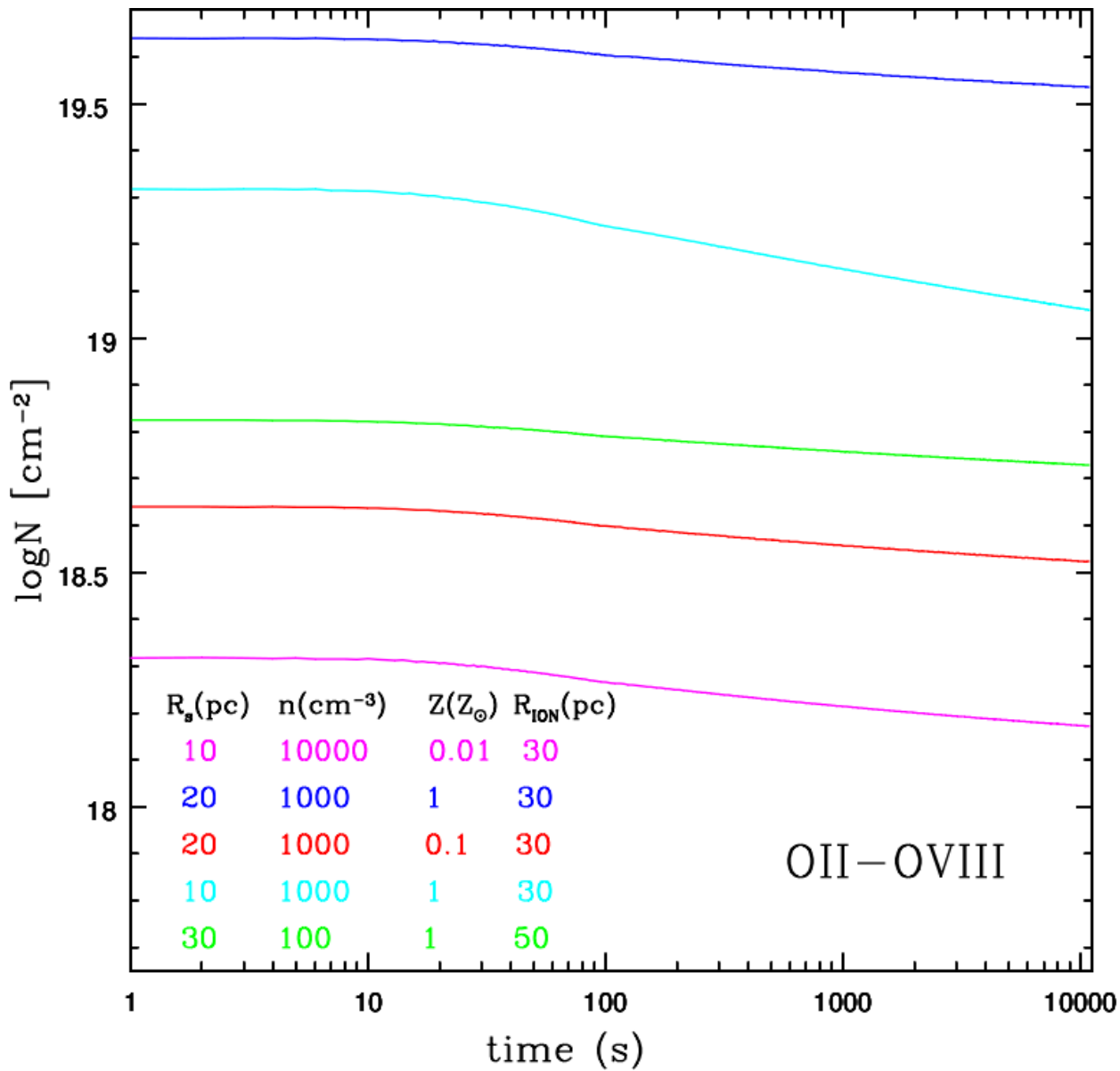} 
\caption[]{Time evolution of the total X-ray column densities produced by
  different step function models of the circumbust
  medium.  Pre-ionization is assumed, with $\rpre=30$\,pc. Only
  integrated column densities for charge states OII to OVIII, as these
  are the ones driving the X-ray opacity. 
  Only mild variations are expected.
 \label{fig:ion_var} } 
\end{figure}


\end{document}